\documentclass[aps,prl,reprint,superscriptaddress,showpacs]{revtex4-2}
\usepackage{graphicx}% Include figure files
\usepackage{dcolumn}% Align table columns on decimal point
\usepackage{bm}% bold math
\usepackage{amsmath}
\usepackage{hyperref}
\usepackage{footnote}
\begin{document}

\title{Limits on the Primordial Black Holes Dark Matter with future MeV detectors}

\author{Zhen Xie}
%\email{zhenxie@mail.ustc.edu.cn}
\affiliation{Deep Space Exploration Laboratory/School of Physical Sciences, University of Science and Technology of China, Hefei 230026, China}
\affiliation{CAS Key Laboratory for Research in Galaxies and Cosmology, Department of Astronomy, School of Physical Sciences,\\
University of Science and Technology of China, Hefei, Anhui 230026, China}
\affiliation{School of Astronomy and Space Science, University of Science and Technology of China, Hefei, Anhui 230026, China}
% ORCID of the author: http://orcid.org/0009-0005-6541-214X

\author{Bing Liu}
\affiliation{Deep Space Exploration Laboratory/School of Physical Sciences, University of Science and Technology of China, Hefei 230026, China}
\affiliation{CAS Key Laboratory for Research in Galaxies and Cosmology, Department of Astronomy, School of Physical Sciences,\\
University of Science and Technology of China, Hefei, Anhui 230026, China}
\affiliation{School of Astronomy and Space Science, University of Science and Technology of China, Hefei, Anhui 230026, China}

\author{Jiahao Liu}
\affiliation{Deep Space Exploration Laboratory/School of Physical Sciences, University of Science and Technology of China, Hefei 230026, China}
\affiliation{CAS Key Laboratory for Research in Galaxies and Cosmology, Department of Astronomy, School of Physical Sciences,\\
University of Science and Technology of China, Hefei, Anhui 230026, China}
\affiliation{School of Astronomy and Space Science, University of Science and Technology of China, Hefei, Anhui 230026, China}

\author{Yi-Fu Cai}
\affiliation{Deep Space Exploration Laboratory/School of Physical Sciences, University of Science and Technology of China, Hefei 230026, China}
\affiliation{CAS Key Laboratory for Research in Galaxies and Cosmology, Department of Astronomy, School of Physical Sciences,\\
University of Science and Technology of China, Hefei, Anhui 230026, China}
\affiliation{School of Astronomy and Space Science, University of Science and Technology of China, Hefei, Anhui 230026, China}

\author{Ruizhi Yang}
\email{yangrz@ustc.edu.cn}
\affiliation{Deep Space Exploration Laboratory/School of Physical Sciences, University of Science and Technology of China, Hefei 230026, China}
\affiliation{CAS Key Laboratory for Research in Galaxies and Cosmology, Department of Astronomy, School of Physical Sciences,\\
University of Science and Technology of China, Hefei, Anhui 230026, China}
\affiliation{School of Astronomy and Space Science, University of Science and Technology of China, Hefei, Anhui 230026, China}

\begin{abstract}

%Primordial black holes (PBHs) emerge as a particularly promising candidate for dark matter, current observations still leave many possible windows in their parameter space.
Primordial black holes (PBHs) are a compelling candidate for Dark Matter (DM). There remain significant parameter spaces to be explored despite  current astrophysical observations have set strong limits.
%This study employs the next-generation MeV observation instrument to constrain the fraction of dark matter attributed to PBHs, denoted as $f_{\rm PBH}$. By observing and estimating the background, we can provide an upper limit on the Hawking radiation of PBHs, and then limit the mass of PBHs in the dark matter halo. 
Utilizing advanced MeV observation instruments, we have statistically established the upper limit of Hawking radiation emitted by PBHs in DM-dense systems, such as galaxy clusters or dwarf galaxies. These results can set a stringent upper limit on the ratio of PBH to DM, expressed as $f_{\rm PBH}$. 
%Our findings demonstrate that conducting MeV observations in Dark Matter (DM) dense regions, such as galaxy clusters or dwarf galaxies, can produce stronger restraints than existing instruments on $f_{\rm PBH}$ in the mass range of $10^{16}-10^{17} ~\rm g$, and also paves a possible way for a novel approach to the observation of PBHs. 
Our results highlight the efficacy of MeV observations in  DM-dense environments. The constraints on $f_{\rm PBH}$ for PBHs in the mass range of $10^{16}-10^{17} ~\rm g$ can be improved significantly compared with the current observations. 

\end{abstract}
\maketitle
\section{Introduction}
A wealth of observational evidence has confirmed that dark matter(DM) constitutes the primary component of the universe\cite{PhysRevLett.125.211101}. However, its preferred model remains an unresolved issue. Primordial black holes (PBHs) are among the earliest proposed and highly motivated candidates for DM \citep{chapline1975cosmological}, potentially formed through the gravitational collapse of high-density regions in the early universe or other exotic mechanisms. Observations of PBHs contribute to our understanding of dark matter by constraining its parameter space and also provide important information on cosmology \citep{Cai:2021zxo,Cai:2021fgm}.
%\textbf{Primordial black holes (PBHs) formed in the early universe proposed in different cosmological scenarios\citep{chapline1975cosmological} are regarded as a compelling candidate for Dark Matter (DM)\citep{Carr_2020}. Observing PBHs can limit the DM parameter space,  and also provide important information on cosmology \citep{Cai:2021zxo,Cai:2021fgm}.}

Despite a wide range of PBHs masses having been excluded by recent observations, the mass range within the $10^{17}-10^{21}\rm g$ remains plausible\cite{Auffinger_2023}.
In addition to the constraints from microlensing and gravitational wave observations, the observations on the Hawking radiations from PBHs \cite{hawking1974black,hawking1975particle} can give direct constraints on the total mass of PBHs in the region of interests (ROIs). %/targets.  observed region. the targets. %
The Hawking radiation of the PBHs in the mass range allowed by current observations peaks at keV to MeV band\cite{10.3389/fspas.2021.681084}. Thus the dedicated astronomical observations in these bands towards the DM-dominated system can provide unique information on PBHs.  However, the MeV observations are severely limited by the sensitivity of current MeV instruments, and the X-ray observations are limited by the small field of view (FOV) and thus expensive exposure time. The planned next-generation MeV detectors will significantly improve the sensitivity and may shed light on the properties of PBHs\cite{ray2021near,coogan2021direct}.

%Given the mass of the single PBH, the total flux of Hawking radiation depends on the total mass of PBHs in the region of interest (ROIs), which is the same as the indirect search of the decaying dark matter\cite{1983PhLB..120..127P}. Thus the best sites for searching the decaying dark matter, such as galaxy clusters and nearby dwarf galaxies, are also ideal sites for searching the PBHs.

%Given the mass of the single PBH, 
The total flux of Hawking radiation is proportional to the total mass of PBHs in the DM halo.  
Thus, same as the indirect search of the decaying DM %\cite{1983PhLB..120..127P}
, the ideal sites for searching the PBHs are nearby dense and massive systems, such as galaxy clusters and dwarf galaxies that in the proximity of our Galaxy.
In this regard, the Perseus galaxy cluster, Perseus for short, renowned as the brightest cluster in the X-ray sky, is one of the most promising sites for such kind of study. The Perseus cluster stands out as a prime candidate for probing CR-induced $\gamma$-ray emissions and DM search \cite{10.1093/mnras/258.1.177,consortium2023prospects}. Its huge mass and proximity could potentially fortify constraints within the PBH mass parameter space. %Consequently, it emerges as a profoundly intriguing target for forthcoming $\gamma$-ray observatories, such as CTA \cite{consortium2023prospects}.

Dwarf spheroidal galaxies, whose mass-to-light ratios reach several hundred in solar units are also regarded as some of the most extensively dark matter-dominated entities in the cosmos. The Draco dwarf spheroidal galaxy in the proximity of our Galaxy, hereinafter referred to as Draco, serves as a representative of this category and was found to be probably DM-dominated by previous studies \cite{lokas2005dark}. Thus, Draco is also an important observational object for next-generation MeV telescopes for constraining the DM theories of PBHs. % validating existing theoretical projections regarding the profiles of DM.
%Extensive research has already been devoted to studying this particular entity.

We therefore chose these two objects to study the possible Hawking radiation of PBHs and constrain the fraction of DM of PBHs ($f_{\rm PBH}$) with hypothetical future MeV detectors. 
The structure of this paper unfolds as follows: In Section 2, we introduce the radiation mechanism of PBHs; in Section 3 we calculate the possible Hawking radiation of PBHs in the Perseus cluster and Draco dwarf galaxy, and then estimate the possible observations using future MeV detectors and derive the expected constraints on $f_{\rm PBH}$ in these two systems;  in the last section discuss the implication of our results.

\section{PRIMORDIAL BLACK HOLES and its Hawking radiation}

The immense compression during the Big Bang could have led to the formation of black holes spanning a diverse range of masses in the early Universe\cite{hawking1971gravitationally}. The elevated density during the early Universe is a crucial factor for PBHs formation, yet it alone is not adequate. It's postulated that substantial primordial irregularities might have existed, allowing overdense regions to halt expansion and undergo subsequent collapse, potentially leading to PBHs formation\cite{nadezhin1978hydrodynamics}.
Alternative theories propose a sort of "designed" inflation where fluctuations' power spectrum—and consequently the production of PBHs reaches its peak at specific scales \cite{hodges1990arbitrariness}.

Given their formation during an era when the Universe was primarily radiation-dominated, these PBHs are considered nonbaryonic, thus eluding the constraints imposed on the baryonic density linked with cosmological nucleosynthesis. Moreover, their current state should be dynamically cold, thereby categorized as one kind of cold DM. Theoretical frameworks proposing the genesis of PBHs during the early universe have sparked discussions suggesting that a significant portion of DM, potentially denoted as $f_{\rm PBH}$, could predominantly consist of PBHs \cite{PhysRevD.94.083504}. 

\begin{figure}[h]
		\centering
        \includegraphics[width=0.45\textwidth]{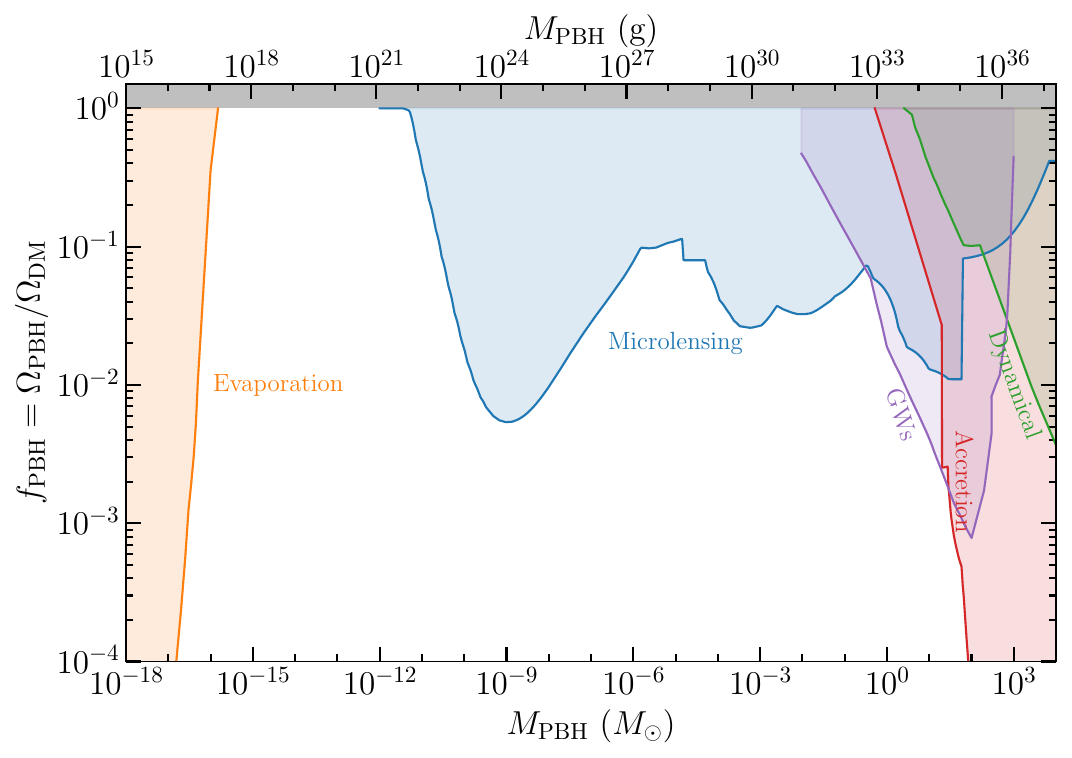} 
        \caption{The constraints from current missions on the fraction of dark matter ($f_{\rm PBH}$) attributed to PBHs vary as a function of PBHs mass ($M_{\rm PBH}$). This figure is crafted using the PBHbounds code \cite{10.5281/zenodo.3538999}.}
	\label{Fig:resultcp}
\end{figure}

PBHs smaller than $10^{15}$\,g should have completely evaporated by now, and the heavier ($\geq 10^{34}$\,g) ones have been excluded by observations \cite{10.3389/fspas.2021.681084}. Fig.\ref{Fig:resultcp} illustrates a compilation of current constraints from various probes, facilitated by the PBHbounds code \cite{10.5281/zenodo.3538999}. The non-observation of microlensing events from the MACHO\cite{allsman2001macho}, EROS\cite{tisserand2007limits}, Kepler\cite{griest2014experimental}, Icarus\cite{oguri2018understanding}, OGLE\cite{niikura2019constraints}, and Subaru-HSC\cite{croon2020subaru}, limits 0.3 solar masses to 30.0 solar masses contribution level. The limitation of PBH evaporation is on the extragalactic $\gamma$-ray background \cite{carr2010new} and on the CMB spectrum \cite{clark2017planck}. And in the heavier mass range, the main limitation comes from searches of stochastic gravitational wave background by LIGO\cite{chen2020distinguishing}, though they could be invalidated\cite{boehm2021eliminating,10.3389/fspas.2021.681084}.

As first shown by Hawking \cite{hawking1974black,hawking1975particle}, a black hole with mass $M = M_{10} \times 10^{10}~\rm g$ emits thermal radiation with temperature:
\begin{align}
    T_{BH}=\frac{1}{8\pi  G M}\sim 1.06 M_{10}^{-1} ~\rm TeV\label{eq1}
\end{align}
and such a black hole emits particles with energy between E and E + dE at a rate:
\begin{align}
    \frac{d^2N}{dEdt}=\frac{1}{2\pi}\frac{\Gamma_s}{e^{E/T_{BH}}-(-1)^{2s}}
\end{align}
where the $\Gamma_s$ is its dimensionless absorption coefficient, the specific form can be obtained from\cite{page1976particle,page1976particle2,page1977particle}.

Particles injected from PBHs exhibit a dual nature comprising two distinct components. The primary constituent arises from direct Hawking radiation, while the secondary element emerges through the decay processes of hadrons formed from the fragmentation of primary quarks and gluons, alongside the decay of gauge bosons, which was first analyzed by MacGibbon and Webber\cite{PhysRevD.41.3052}. As an illustration, the comprehensive photon spectrum emitted by a low-mass PBH can be expressed as follows:
\begin{align}
    \frac{d^2N}{dE_\gamma dt}(E_\gamma,M)=\frac{d^2N}{dE_\gamma dt}(E^{pri}_\gamma,M)+\frac{d^2N}{dE_\gamma dt}(E^{sec}_\gamma,M)
\end{align}

A remaining window allowed by current observations persists within the mass range of $10^{16}-10^{21}\rm g$. The PBHs in this mass range can emit Hawking radiation in the keV-MeV band, which presents a possibility of constraining this parameter space through observations in hard X-ray to gamma-rays. For instance, positive outcomes have been deduced in the electron-positron (511 keV) annihilation line observations near the Galactic Center \cite{PhysRevLett.123.251101,derocco2019constraining,dasgupta2020neutrino}, the CMB power spectrum \cite{Acharya_2020}, and measurements of 21-cm signal distortion \cite{PhysRevD.105.103026,Mittal_2022}. Additionally, X-ray surveys conducted within the inner regions of the Milky Way or dwarf spheroidal galaxies have contributed to this exploration \cite{malyshev2022search,Auffinger_2023}.

\begin{figure}[h]
		\centering
        \includegraphics[width=0.45\textwidth]{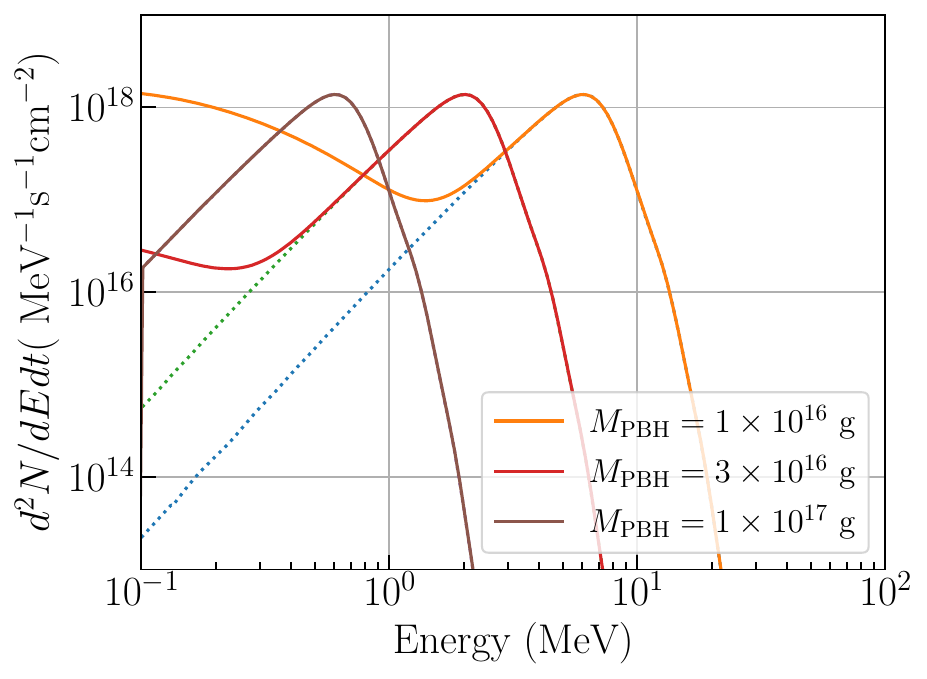} 
        \caption{The spectrum of PBHs with different mass, calculated by {\it BLACKHAWK }v2.2. The dashed line shows the primary emitted Hawking radiation. }
	\label{Fig:pbhm}
\end{figure}

To determine the spectrum of PBHs, we employ the publicly available software {\it BLACKHAWK }v2.2\cite{arbey2019blackhawk,arbey2021physics}. This software facilitates the calculation of secondary particle generation arising from various processes like hadronization, fragmentation, decay, and other processes of black hole evaporation. The secondary spectra are contingent on the development of Standard Model particles released through Hawking radiation. {\it BLACKHAWK} includes public codes like \textit{PYTHIA}\cite{sjostrand2015introduction} and \textit{HERWIG}\cite{bellm2020herwig} to calculate these process. Fig.\ref{Fig:pbhm} shows the photon spectra from the Hawking radiation from a single PBH with the mass of $10^{16}~\rm g$, $3\times 10^{16}~\rm g$ and $10^{17}~\rm g$, respectively.

\section{Bounds on $f_{\rm PBH}$ with MeV observations}
As introduced in Section 1, we chose Perseus and Draco as our objects of study to investigate the possible observations in the MeV band.  We consider the halo to fulfill the spherical collapse model for overdensities $\Delta = 200$ times the critical density of the Universe $\rho_{crit}$. The radius can be calculated by
\begin{align}
    R_{200}=(\frac{M_{200}}{\frac{4}{3}\pi \Delta\rho_{crit}}),\theta_{200}=arctan(\frac{R_{200}}{d_L})
\end{align}
For Perseus, $M_{200}=7.5\times 10^{14} ~\rm M_\odot$,$R_{200}=1.865~\rm Mpc$\cite{consortium2023prospects}.  For Draco we used  $M_{200}=1.8\pm0.7\times 10^9 ~\rm M_\odot$\cite{read2019dark}.  The angular size of the DM halo, $\theta_{200}$, can be estimated respectively as $1.4^{\circ}$ and  $1.3^{\circ}$ for Perseus and Draco, assuming the distance of these two objects is 75 Mpc \cite{read2019dark} and 80 kpc \cite{consortium2023prospects}. 

We first consider the case where the PBH mass distribution is monochromatic. Employing {\it BLACKHAWK}, we compute the photon spectrum $F_{\rm PBH}(M_{\rm PBH})$ spanning from 1 MeV to 100 MeV for single PBH with each specific mass $M_{\rm PBH}$. In this case, the total Hawking radiation from the DM halo relates only to the total mass of the halo, rather than the specific DM spatial profile. Thus the total photon flux from the PBHs in Draco and Perseus can be easily estimated as $F_{\rm total}= f_{\rm PBH} (M_{200}/M_{\rm PBH}) F_{\rm PBH}$ for given $f_{\rm PBH}$.  

The statistical counts recorded by instruments adhere to the Poisson distribution, enabling us to establish an upper limit of $F_{\rm total}$ and thus $f_{\rm PBH}$  for estimating the unobservable outcome based on background counts. Determining the anticipated background involves integrating the instrument's effective area, the background spectrum, and the observation period. 

Recently,  many projects for next-generation MeV detectors have been proposed such as e-ASTROGAM \citep{de2018science}, AMEGO\citep{2019BAAS...51g.245M},  COSI\citep{tomsick2019compton} and  MeGaT \cite{megat}, which all reveal a significant improvement in the sensitivities compared with the current MeV instruments. Rather than using the instrumental response of specific instruments, we assume the next generation MeV instruments have an effective area of $100~\rm cm^2$ and a point spread function (PSF) of $2^\circ$, which is reasonable taken into the design and preliminary simulation results for both the semiconductor detectors such as e-ASTROGAM and gas detectors such as MeGaT. 

In contrast to observations in the X-ray energy range, which typically occur over kiloseconds, we consider a longer timescale for MeV observations. This extended observation time is justifiable as the planned MeV detectors all have a much larger field of view and do not necessitate pinpointing a specific point source for observation. Thus the observation time of the MeV telescope is not as expensive as that of X-rays. In this work, we chose a length of two months to calculate the background counts.

In addition to the heavy mass, Perseus and Draco lie both in high galactic latitude, which makes them suffer much less contamination from other bright $\gamma$-ray emitters and the diffuse Galactic background in the MeV band, thus significantly improving the sensitivity of detecting/constraining the radiations from DM. It should be noted that in the calculation the PSF of the instrument is larger than the angular size $\theta_{200}$ for both Perseus and Draco, thus we calculated the background emission based on the instrumental PSF ($2^{\circ}$ for searching the PBH signal. 
\begin{figure}[h]
		\centering
        \includegraphics[width=0.4\textwidth]{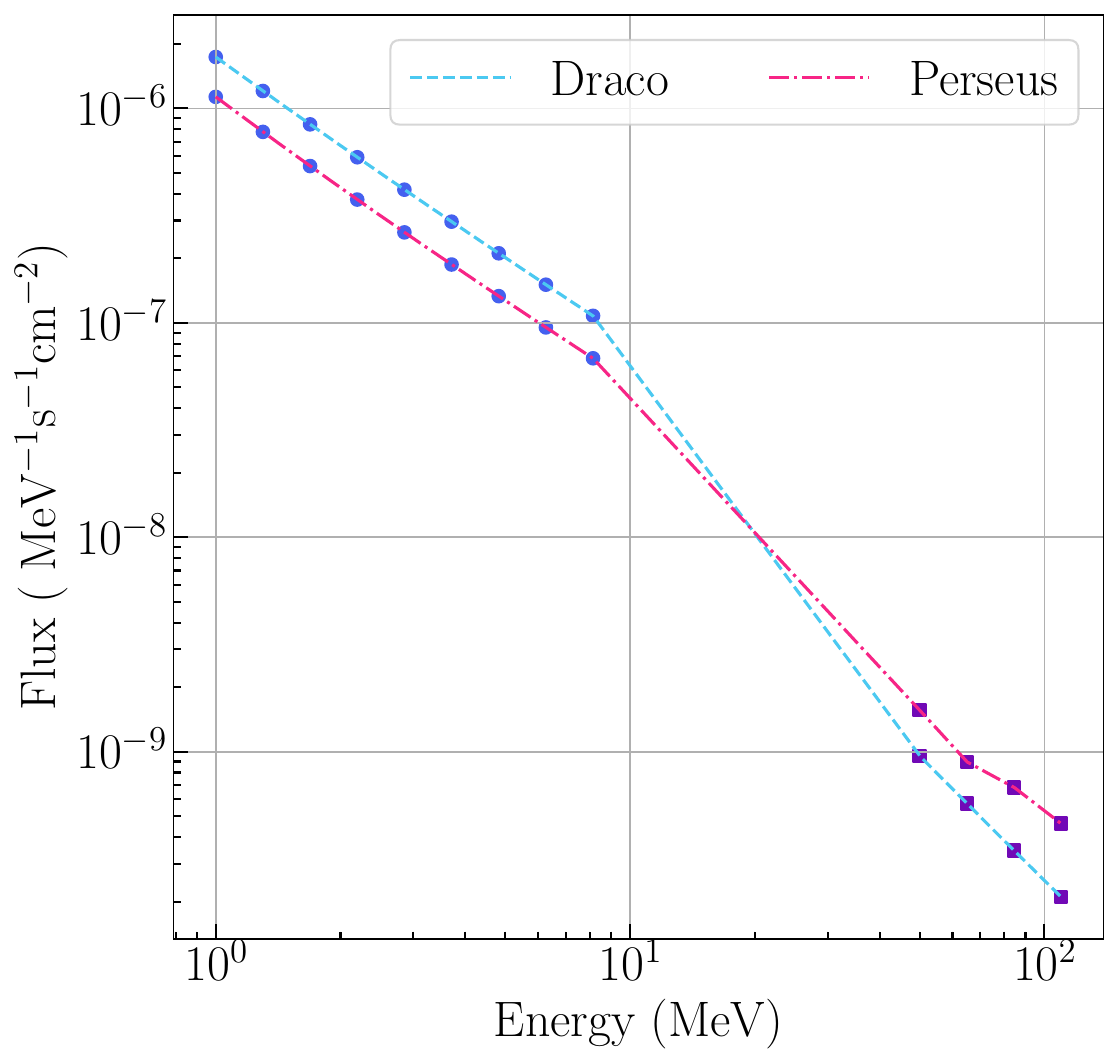} 
        \caption{Background flux in Perseus and Draco region with a radius of 2 degrees, the background of 1-10 MeV is extrapolated from the results of \cite{Siegert_2022}, and 50-100 MeV is from the work of Fermi-LAT\cite{fermibkg}. The dashed line is the function we use to calculate the background by interpolating these data. }
	\label{Fig:bkg}
\end{figure}

The diffuse emissions below $10~\rm MeV$ around the Galactic center were recently reanalyzed by \citet{Siegert_2022} using Integral SPI observations. For a rough estimation,  we extrapolate the results to the position of Draco and Perseus assuming the spatial distribution of the MeV band background can be described by the same energy-dependent spatial template predicted by GALPROP models \citep{galprop} that were used in \citet{Siegert_2022}. %({\bf add some more detail on this galprop model version}) \citep{Siegert_2022}. 
%And the background of 50-100 MeV is given by the official diffuse background model of Fermi-LAT\cite{fermibkg}.  
For the diffuse background over 50MeV, we adopted the interstellar emission model\citep{fermibkg} based on the first 9 years of Fermi-LAT science data, {\it gll\_iem\_v07.fits}, and integrated the flux from the direction of Draco and Perseus within 2 degrees. We connect the two values smoothly by interpolation in the energy range between $10~\rm MeV$ and $50~\rm MeV$. The results for both Perseus and Draco (for a ROI/PSF of $2^{\circ}$) are shown in Fig.\ref{Fig:bkg}.  

\begin{figure}[h]
	\centering
        \includegraphics[width=0.4\textwidth]{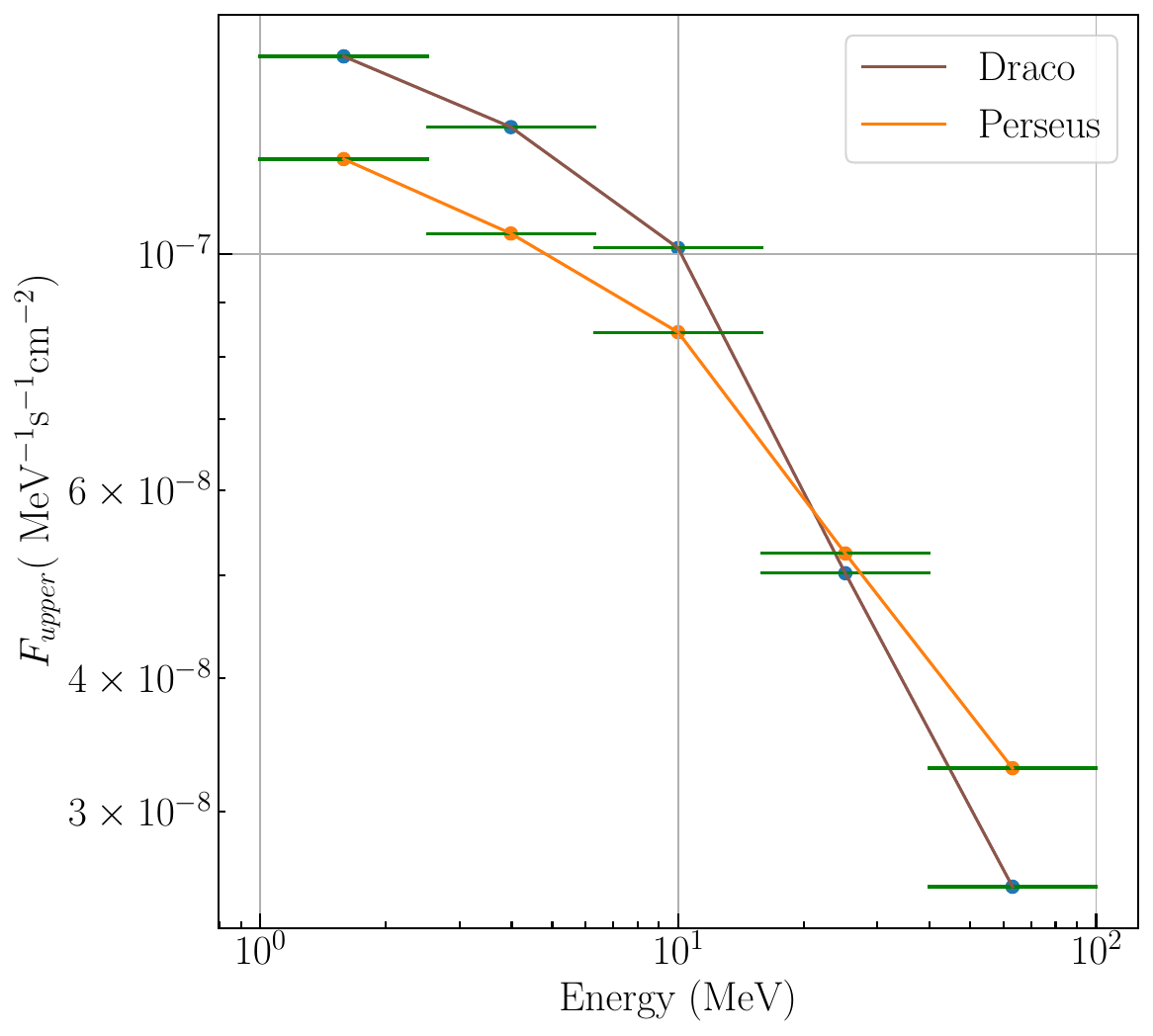} 
        \caption{The sensitivity (or upper limit, in the nondetection case) of flux from PBH assuming a ROI/PSF of $2^{\circ}$, expressed as$F_{upper}$, in the Draco and Perseus region. The green bar represents the integral interval of the energy component bin.}
	\label{Fig:fup}
\end{figure}

Given the background flux estimated above, the background photon counts $N_{bkg}$ can be estimated by multiplying the flux with the effective area $A_{\rm eff}$ and the exposure time $T_{\rm exp}$. We calculate the photon counts by dividing the 1-100 MeV energy interval into five uniform bins in logarithmic space. Photon counts follow the Poisson distribution, and the $3\sigma$ fluctuation can be estimated  $ N_{3\sigma} = 3\times \sqrt{N_{\rm bkg}}$. In a simple on-off analysis, the $ N_{3\sigma}$ can then be used to estimate the sensitivity of the detection. Thus the sensitivity (or upper limit, in the non-detection case) of flux from PBH can be estimated as $F_{upper} = N_{3\sigma}/(A_{\rm eff}T_{\rm exp})$, shown in Fig. \ref{Fig:fup}. The derived $F_{\rm upper}$  can then be compared with the theoretical expected $F_{\rm total}$ for each PBH mass $M_{\rm PBH}$ to derive the upper limit on $f_{\rm PBH}$.  

\begin{figure}[hp]
		\centering
        \includegraphics[width=0.45\textwidth]{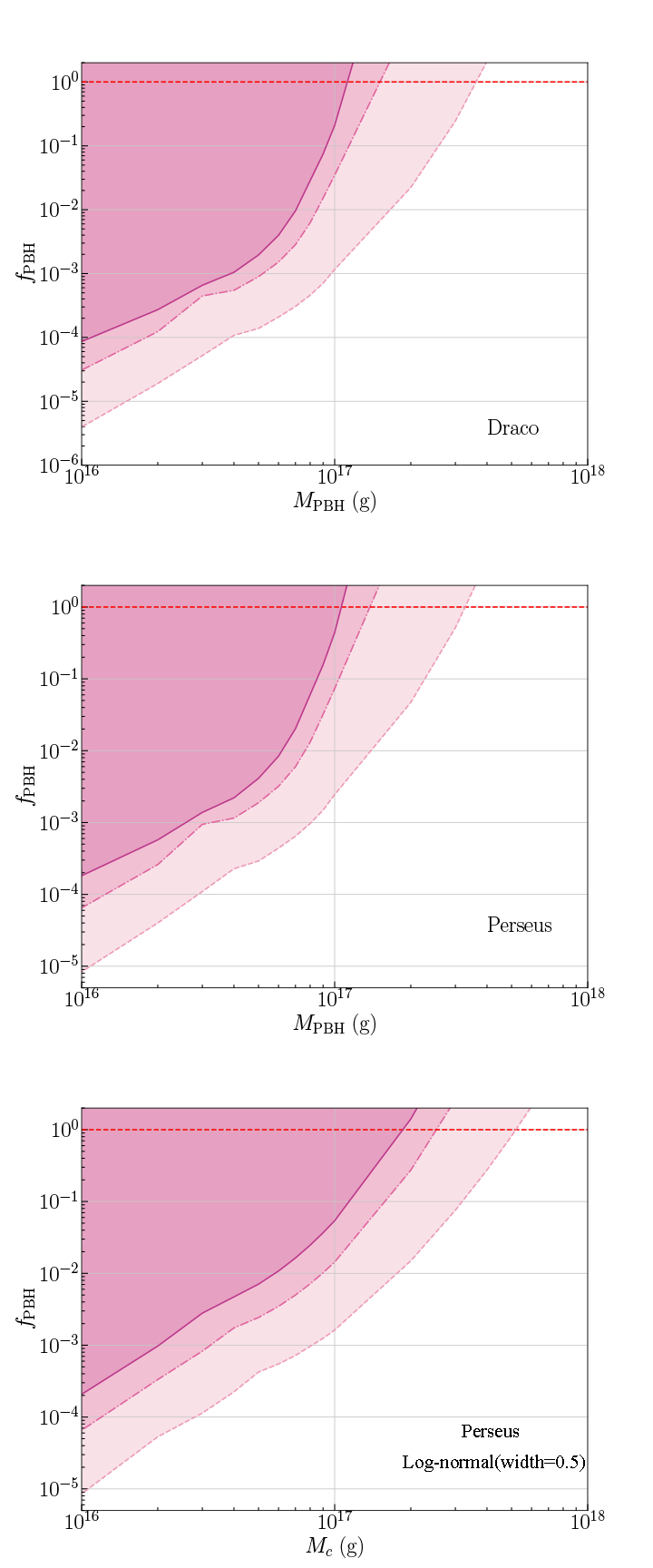} 
        \caption{The 3$\sigma$ limitation that MeV observation of Perseus and Draco can provide for 2 months time length. Different depths of color show different reduced spin parameters, solid,dot-dashed, and dashed lines correspond to $a_*=0, 0.5$ and $0.9$. The bottom picture shows the result of a Log-normal distribution of mass for $\sigma=0.5$ with change of different peak mass $M_c$.}
	\label{Fig:realts}
\end{figure}

To consider the effect of PBHs spin, we choose different values of the reduced spin parameter $a_*=J/(G_NM^2_{PBH})=0,0.5,0.9$, where $J$ is the angular momentum of the PBH. Theoretically, the spinning BHs evaporate faster\cite{page1976particle,page1976particle2,page1977particle}, so with a larger value of $a_*$, the radiation is stronger than the nonspinning cases, also the limitation will be more stringent.
The results for the bounds on $f_{\rm PBH}$ are shown in Fig.\ref{Fig:realts}. 

If the PBHs span an extended range of masses, the mass function is usually written as $dn/dM$. The distribution of BHs as a function of their mass or any other parameter (spin, charge) is completely model-dependent. To illustrate the effect, here we consider the Log-normal 
distributions\cite{PhysRevD.47.4244}
\begin{align}
    \frac{dn}{dM}=\frac{A}{\sqrt{2\pi}\sigma M}exp(-\frac{ln(M/M_c)^2}{2\sigma^2})
\end{align}
where A is the amplitude, $M_c$ is the position of the peak and $\sigma$ is its width. This model is a good approximation if the PBHs are formed from a smooth symmetric peak in the inflationary power spectrum\cite{PhysRevD.96.023514}. We take $\sigma=0.5$ as an example to show the effect of the mass distribution function, and the results are also presented in Fig.\ref{Fig:realts}. In this case the limit on $f_{PBH}$ is also slightly more stringent than in the monochromatic mass case. 

\begin{figure}[h]
		\centering
        \includegraphics[width=0.45\textwidth]{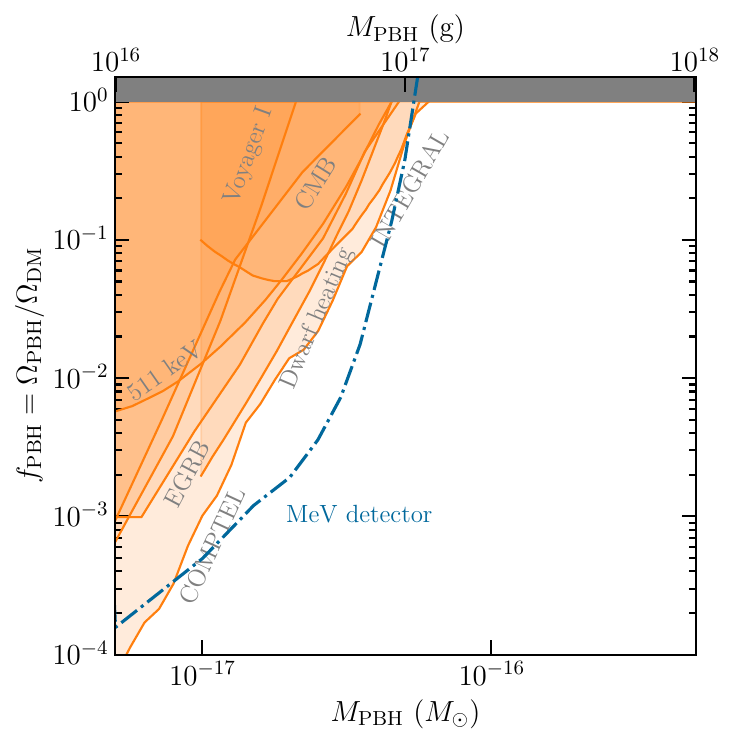} 
        \caption{The constraints within the PBHs mass range of $10^{16}-10^{18}\rm g$. Our results of Perseus, for the non-spin monochromatic mass situation, represented by the cerulean line, encapsulate a two-month observation period utilizing MeV detectors. This figure is crafted using the PBHbounds code \cite{10.5281/zenodo.3538999}.}
	\label{Fig:resultevcp}
\end{figure}

The result of Draco is very close to what we got in Perseus, this is due to the distance of Draco (80 kpc) being much closer than Perseus (75 Mpc), and the background levels are very similar in these two regions as shown in\ref{Fig:bkg}. We note the total mass of the DM halo in Draco has nonnegligible uncertainties, which is shown as the shaded area in Fig.\ref{Fig:realts}.  We also compared our result of Perseus with existing limitations, as shown in Fig.\ref{Fig:resultevcp}. We found that our results can improve significantly for the mass range between $10^{16}~\rm g$ and $10^{17}~\rm g$.  

\section{Discussion}
In this article, we present the possible observation of Hawking radiation from PBHs based on the performance of next-generation MeV detectors. We found that in the DM halo of both galaxy cluster and dwarf spherical galaxy, the MeV observations with reasonable exposure can significantly improve the current constraints of PBHs as DM, especially in the PBHs mass range of $10^{16}~\rm  g$ to  $10^{17}~\rm g$. 

It should be noticed that in this range of mass, 21 cm observations, using its temperature\cite{saha2022sensitivities} and redshift information\cite{mittal2022constraining}, can give us more stringent constraints. However, these bounds are highly model-dependent, and have not yet been confirmed by other surveys, a more objective approach is to put all of these limits as prospective\cite{auffinger2023primordial}.  In comparison, the constraints from Hawking radiation depend only on the total mass of the DM halo, which is well established by the observations of star/galaxy dynamics. Thus the constraints we expected here are robust and nearly model-independent. 

As shown in \cite{malyshev2023limits} the current and future X-ray observations can also improve the constraints on $f_{\rm PBH}$. Indeed, as shown in Eq.\ref{eq1} the X-ray and MeV bands are most suitable for different PBH mass ranges. Furthermore, the X-ray instruments, limited by their much smaller field of view, can be difficult to observe such large structures ($\theta_{200}$ are more than 1 degree for both Perseus and Draco), and the long exposure on such regions can also be very expensive. In this regard, the large FOV MeV instruments are more suitable for such kind of study. 

In conclusion, the future MeV observations show promising prospects in constraining the PBHs as DM candidates. It should be noted that the methodology here can also be used to search for other decaying DM candidates. Due to the lack of sensitivity of current MeV instruments, the parameter space of particle DM candidates in the mass range below several GeV is largely unexplored, and future MeV projects would shed light on the indirect search of DM.  
\section{acknowledgment} 
R. Yang is supported by the NSFC under grant 12041305 and the national youth thousand talents program in China. Bing Liu acknowledges the support from the NSFC under grant 12103049.
\appendix

\bibliography{main.bib}

\end{document}